\documentclass[11pt]{article}
\usepackage{amssymb}
\usepackage{amsmath}
\usepackage{amscd}
\usepackage{latexsym}
\usepackage{graphicx}
\usepackage{here}
\usepackage{bbm}
\usepackage{psfrag}

\catcode`@=11
\renewcommand{\section}{\@startsection{section}{1}{0pt}{\medskipamount}
{\medskipamount}{\large\bf}}
\numberwithin{equation}{section}
\catcode`@=12

\def\a{\alpha}
\def\b{\beta}
\def\ab{\bar{\alpha}}
\def\zb{\bar{z}}
\def\at{\widetilde{\alpha}}
\def\ph{\widehat{\phi}}

\def\sfrac#1#2{{\textstyle\frac{#1}{#2}}}
\def\+{\dagger}
\def\={\ =\ }
\def\pa{\partial}

\def\>{\rangle}
\def\<{\langle}

\newcommand{\unity}{\mathbbm{1}}
\newcommand{\e}{\,\mathrm{e}\,}
\newcommand{\im}{\,\mathrm{i}\,}
\newcommand{\diff}{\mathrm{d}}
\newcommand{\tr}{\mathrm{tr}}

\newcommand{\R}{{\mathbb{R}}}
\newcommand{\C}{{\mathbb{C}}}
\newcommand{\Z}{{\mathbb{Z}}}
\newcommand{\NN}{{\mathbb{N}}}

\newcommand{\be}{\begin{equation}}
\newcommand{\ee}{\end{equation}}
\newcommand{\bea}{\begin{eqnarray}}
\newcommand{\eea}{\end{eqnarray}}
\newcommand{\bal}{\begin{aligned}}
\newcommand{\eal}{\end{aligned}}

\newcommand{\und}{\qquad\textrm{and}\qquad}

\begin{document}

%%%%%%%%%%%%%%%%%%%%%%%%%%%%%%%%%%%%%%%%%%%%%%%%%%%%%%%%%%%%%%%%%%%%%%

\begin{titlepage}
\setcounter{page}{0}
\begin{flushright}
hep-th/0604219\\
ITP--UH--08/06\\
\end{flushright}
\vskip 2.0cm

\begin{center}

{\LARGE\bf 
Moduli-Space Dynamics of \\[12pt]
Noncommutative Abelian Sigma-Model Solitons}

\vspace{15mm}

{\Large
Michael Klawunn , \ Olaf Lechtenfeld \ and \ Stefan Petersen }
\\[10mm]
\noindent {\em Institut f\"ur Theoretische Physik,
Universit\"at Hannover \\
Appelstra\ss{}e 2, 30167 Hannover, Germany }\\
{Email: klawunn, lechtenf, petersen @itp.uni-hannover.de}
\vspace{20mm}

\begin{abstract}
\noindent
In the noncommutative (Moyal) plane, we relate exact U(1)~sigma-model 
solitons to generic scalar-field solitons for an infinitely stiff potential.
The static $k$-lump moduli space $\C^k/S_k$ features a natural K\"ahler metric
induced from an embedding Grassmannian. The moduli-space dynamics is blind
against adding a WZW-like term to the sigma-model action and thus also applies
to the integrable U(1)~Ward model. For the latter's two-soliton motion we 
compare the exact field configurations with their supposed moduli-space 
approximations. Surprisingly, the two do not match, which questions the 
adiabatic method for noncommutative solitons.
\end{abstract}

\end{center}
\end{titlepage}

\section{Introduction and summary}
\noindent
Field theories with a nontrivial vacuum structure often feature 
static localized finite-energy solutions. Such lumps can be boosted
to single solitons moving with constant velocity. The scattering of
these solitons off one another is, however, usually accessible only 
numerically.\footnote{
Exceptions are integrable theories, which allow for analytic
multi-soliton configurations and an exact S-matrix.}
Alternatively, a qualitative understanding of soliton scattering can
be achieved for small relative velocity via the adiabatic or moduli-space
dynamics invented by Manton~\cite{manton,mantonbook}. 
This approach approximates the exact $k$-soliton scattering configuration by 
a time sequence of static $k$-lump solutions. Thereby one introduces a time
dependence for the latter's moduli~$\alpha_i$, which is determined by 
extremizing the action on the moduli space~${\cal M}_k$. Being a functional 
of finitely many moduli~$\alpha_i(t)$, this action describes the motion 
of a point particle in~${\cal M}_k$, equipped with a metric~$g_{ij}(\alpha)$
and a magnetic field~$A_i(\alpha)$. Hence, the motion of $k$ slowly scattering 
solitons is well described by a geodesic trajectory in~${\cal M}_k$,
possibly with magnetic forcing. Since among the moduli are the spatial 
locations of the individual quasi-static lumps, (a projection of)
the geodesic in~${\cal M}_k$ may be viewed as trajectories of the various 
lumps in the common ambient space.

In order to test the validity of the adiabatic method, one would need to 
apply it to an integrable model, where exact multi-soliton solutions are
available for comparison. Yet, such theories are rare in two or more
spatial dimensions, which are required for interesting trajectories.
A prime example is the nonlinear sigma model for some group~${\cal G}$ in 
$1{+}2$ dimensions. By adding a (necessarily Lorentz-breaking) WZW-like term
with an arbitrary coefficient, one generates a one-parameter family of 
extended sigma models. Their common static configuration space is well known
to contain multiple lumps (one-unitons), which are based on hermitian 
projectors and thus sit in certain Grassmannians~\cite{uhl,wood,zakbook}.
For a particular strength of the WZW-like term, one obtains the Ward model,
which is integrable~\cite{ward1,ward2,ioazak}. Recently, the adiabatic 
approach was tested in this model for ${\cal G}=$SU(2)~\cite{dunman}.

It is rewarding to generalize the above set of ideas to field theories on
noncommutative spaces. Such field theories offer not only smooth deformations
of well known soliton solutions but entirely new types of noncommutative
solitons (for lectures on the subject see~\cite{komaba}). 
This property is most prominent when the commutative limit 
yields a free theory, because the soliton configurations are then forced 
to become singular in this limit. A case in point is the abelian sigma model, 
i.e.~choosing ${\cal G}=$U(1), on a noncommutative plane with ordinary time. 
It has the virtue that its static $k$-lump solutions take a very simple form
and depend exclusively on the $k$ complex location moduli~\cite{lepo1,dolepe}. 
Furthermore, its extension \`a la Ward is again integrable~\cite{lepospe}. 
Therefore, the family of extended U(1)~sigma models seems ideally suited to try
out the adiabatic method in the noncommutative realm, and this is what we do 
in the present paper. For definiteness, we work with the standard Moyal
deformation, labelled by a positive constant parameter~$\theta$. 
The Moyal-Weyl map is employed to pass from the star product to the operator 
realization on the harmonic-oscillator Fock space~${\cal H}$.

The sigma-model constraint can be implemented in unconstrained
(multi-component) scalar field models by choosing an appropriate potential
and performing an infinite-stiffness limit. Therefore, the soliton analysis 
for generic noncommutative field theories in $1{+}2$ dimensions
\cite{komaba,gomistr,agomistr,liroun,goheaspr,haliroun,haimoh} 
applies to noncommutative sigma models (without WZW-like term) as well. 
In fact, it yields the exact static multi-lumps on the Moyal plane for 
any value of~$\theta$, as we shall review in section~2 below.
More precisely, the $k$-lump moduli space~${\cal M}_k$ is parametrized by
all collections of $k$~harmonic-oscillator coherent states and is a
$k$-dimensional complex submanifold of the Grassmannian Gr($k,{\cal H}$).
Section~3 computes the full moduli-space action for the noncommutative U(1)
extended sigma model. For reasons to be explained, the result turns out 
to agree with the $\theta$-independent part of the moduli-space action for 
the generic scalar field theory, irrespective of the WZW-like term.
More concretely, ${\cal M}_k$ has a natural K\"ahler structure, with the 
K\"ahler potential being given by the determinant of the matrix of 
coherent-state overlaps~\cite{goheaspr}. There is no magnetic background field.
We briefly discuss the properties of the moduli-space metric 
and its limits for coinciding lumps.

The moduli-space scattering trajectories for two solitons are investigated 
in section~4 and exhibit scattering angles between $0$ and~$\frac\pi2$. 
Finally, section~5 compares with the time-dependent solutions of the field 
theory family, in particular with the exact multi-soliton configurations 
of the U(1)~Ward model. Barring some miracle, the latter features only
solitonic no-scattering or bound-state solutions, which we display.
It appears that the adiabatic approximation fails in this integrable case, 
possibly due to its inability to sense the WZW-like action term.
A numerical investigation, also away from the integrable case, could help
settle this issue. Finally, we remark that our considerations are purely 
classical and likely to be modified by quantum corrections.

\bigskip

\section{Noncommutative static lumps in scalar field and sigma models}
\noindent
We begin with a rather generic action of a real scalar field~$\phi$ living
on the Moyal plane with coordinates~$(z,\bar z)$ and depending on time~$t$,
\be
S_\theta \= \int\!\diff{t}\,\diff^2{z}\; \bigl[ \sfrac12 \dot\phi^2 
- \partial_z\phi\,\partial_{\bar z}\phi - V_\star(\phi) \bigr] \quad,
\ee
where the subscript on the potential signifies star-product multiplication
based on 
\be
z \star \bar z - \bar z \star z = 2\,\theta \quad.
\ee
We further specify
\be
V(\phi)\ge 0 \quad,\qquad
V(\phi_0)=0 \qquad\textrm{and}\qquad
V'(\phi) \= v\prod_i (\phi-\phi_i) \quad.
\ee
Static classical configurations~$\phi_{\textrm{cl}}$ 
extremize the energy functional
\be
E_\theta \= \int\!\diff^2{z}\; \bigl[
\partial_z\phi\,\partial_{\bar z}\phi + V_\star(\phi) \bigr] \quad,
\ee
which for large values of~$\theta$ is dominated by the potential term,
because $z=O(\sqrt{\theta})$.
Expanding around $\theta{=}\infty$, one obtains~\cite{goheaspr}
\be \label{phicl}
\phi_{\textrm{cl}} \= \ph + \sfrac{1}{\theta} \tilde\phi +\ldots
\qquad\textrm{with}\qquad
\ph \= \sum_i \phi_i\,P_i \quad,
\ee
where $\{P_i\}$ is an orthogonal resolution of the star-algebra identity,
\be
P_i\star\!P_j \= \delta_{ij}P_j 
\qquad\textrm{and}\qquad \sum_i P_i \= \unity \quad.
\ee
We also introduce the rank~$k_i$ of~$P_i$ via
\be \label{rankdef}
\int\!\sfrac{\diff^2{z}}{2\pi\theta}\;P_i \= k_i \in \NN_0 \quad.
\ee
Since $\tilde\phi$ and all further terms in the expansion are determined 
by~$\ph$, any classical solution is fixed by an assignment of projectors~$P_i$
to the extrema~$\phi_i$ of the potential. Restricting ourselves to stable 
solutions, we always associate the zero projector (which is admissible) to the
local maxima of~$V$. Please note that the collection~$\{P_i\}$ appearing 
in~(\ref{phicl}) must be complete. The expansion of the classical energy reads
\be
E_\theta [\phi_{\textrm{cl}}] \= \theta E_0+E_1+\sfrac{1}{\theta}E_2 +\ldots
\qquad\textrm{with}\qquad
E_0 \= 2\pi\,\sum_i k_i\ V(\phi_i)
\ee
and
\be \label{E1}
E_1 \= \int\!\diff^2{z}\ \partial_z\ph\,\partial_{\bar z}\ph
\= \sum_{ij} \phi_i\,\phi_j \int\!\diff^2{z}\ 
\partial_z P_i\,\partial_{\bar z} P_j \quad.
\ee
Any complete collection~$\{P_i\}$ extremizes $E_\theta$ at leading order 
in~$\theta$. Beyond this, however, $E_1$ lifts this infinite degeneracy:
its extremization selects a finite-dimensional class of identity resolutions.

In \cite{goheaspr}, an asymmetric double-well potential was chosen,
with local minima $V(0)=0$ and $V(\lambda)>0$. The authors assigned
\be
\phi=\lambda \quad\leftrightarrow\quad P \und
\phi=0 \quad\leftrightarrow\quad \unity{-}P \quad,
\ee
which led to
\be
\ph \= \lambda\,P \qquad\textrm{as well as}\qquad 
E_0 \= 2\pi\,k\,V(\lambda) \und 
E_1 \= \lambda^2 \smallint\!\diff^2{z}\;|\partial_z P|^2 \quad.
\ee
Presently, we consider instead the double-well potential
\be
V(\phi) \= \beta\,(\phi^2-1)^2  \qquad\longrightarrow\qquad
V'(\phi) \= 4\beta\,(\phi{+}1)\,\phi\,(\phi{-}1)
\ee
and associate
\be
\phi=-1 \quad\leftrightarrow\quad P \und
\phi=+1 \quad\leftrightarrow\quad \unity{-}P \quad,
\ee
which implies 
\be \label{phihat}
\ph \= \unity - 2P \qquad\textrm{as well as}\qquad
E_0 \= 0 \und
E_1 \= 4 \int\!\diff^2{z}\;|\partial_z P|^2 \quad.
\ee
It is easy to see that all higher corrections, i.e.~$\tilde\phi$, $E_2$ etc.,
come with negative powers of~$\beta$. Therefore, we have the exact result
\be
\phi_{\textrm{cl}}\ \to\ \ph \und 
E_\theta [\phi_{\textrm{cl}}]\ \to\ E_1
\ee
in the limit of infinite stiffness, $\beta\to\infty$, 
and there is no effective potential on the moduli space. 
This limit nails the value of~$\phi$ to $-1$ (in im$P$) or to $+1$ (in ker$P$)
and makes the classical configuration idempotent, i.e.~$\ph^2_\star=\unity$.

Idempotent fields also appear in nonlinear sigma models, where they define
Grassmannian submanifolds of the group~${\cal G}$ via $P=\sfrac12(1{-}\phi)$. 
The simplest case occurs for ${\cal G}=$U(1), 
i.e.~for complex unimodular~$\phi$, 
and becomes interacting when being Moyal deformed.\footnote{
The action will be given in the following section.}
Its two Grassmannian submanifolds correspond precisely to 
the two idempotent values above, namely $\phi=\pm1$. Hence, 
if we extend our double-well model to a Mexican-hat model for complex~$\phi$,
the stiff limit will yield the constraint~$|\phi|_\star^2=\unity$ defining
the $\textrm{U}_\star(1)$ sigma model, and our trial configurations~$\ph$
for $P$ of rank~$k$ parametrize precisely the Grassmannian Gr($k,{\cal H}$).
The only modification owed to the extension is a factor of two in the energy
functional. For $k<\infty$, the latter can be manipulated to
\be
E_1 \= 8 \int\!\diff^2{z}\ |\partial_z P|^2 
\= 8\pi\,k\ +\ 16 \int\!\diff^2{z}\ 
|(\unity{-}P)\star\partial_{\bar z}P|^2 \ \ge\ 8\pi\,k \quad,
\ee
revealing a Bogomolnyi bound.\footnote{
A finite~$k$ also agrees with the topological charge~$Q$ of the respective 
Grassmannian. Negative values of~$Q$ are produced by the flip 
$P\longleftrightarrow\unity{-}P$ and correspond to anti-solitons.}
The saturation $E_\textrm{BPS}=8\pi k$ is reached when
\be \label{BPS1}
(\unity{-}P)\star\partial_{\bar z}P\=0 \quad,\qquad\textrm{i.e.}\quad
\partial_{\bar z}\!: \textrm{im}P \hookrightarrow \textrm{im}P \quad.
\ee
Hence, the static classical configurations~$\ph=\unity{-}2P$ extremizing~$E_1$
are given by projectors stable under the $\partial_{\bar z}$ action,
and their moduli space~${\cal M}_k$ for rank~$k$ describes static $k$-lump 
solutions of the $\textrm{U}_\star(1)$ sigma model in the Moyal plane.

In order to find an explicit parametrization of~${\cal M}_k$, we pass from
the star-product to the operator formulation,
\be
z\ \rightarrow\ \sqrt{2\theta}\,a \quad,\qquad
\bar z \ \rightarrow\ \sqrt{2\theta}\,a^\+ \qquad\textrm{hence}\qquad
\sqrt{2\theta}\,\partial_z \ \rightarrow\ -[a^\+,.] \quad,\qquad
\sqrt{2\theta}\,\partial_{\bar z}\ \rightarrow\ [a,.]
\ee
with
\be
[a\,,\,a^\+]\=\unity  \und
\smallint\!\diff^2z\;\dots \= 2\pi\theta\;\tr_{\cal H} \dots \quad,
\ee
justifying the definition~(\ref{rankdef}).
The Fock space~${\cal H}$ representing this Heisenberg algebra is spanned
by the basis
\be \label{nbasis}
|n\> \= \sfrac{1}{\sqrt{n!}}\,(a^\+)^{n}\,|0\> \qquad\textrm{with}\qquad 
n\in\NN_0 \und a\,|0\>=0 \quad.
\ee
Any rank-$k$ projector in~${\cal H}$ can be decomposed as
\be
P \= |T\> \sfrac{1}{\<T|T\>} \<T| \qquad\textrm{with}\qquad
|T\> \= \bigl( |T_1\>, |T_2\>, \dots, |T_k\> \bigr) \quad,
\ee
where the (not necessarily orthonormal) states $|T_i\>$ 
span the image of~$P$. The BPS equation~(\ref{BPS1}) now reads
\be \label{BPS2}
(\unity{-}P)\,a\,P \=0  \qquad\Longleftrightarrow\qquad
a\,|T\> \= |T\>\,\Gamma 
\qquad\textrm{for some $k{\times}k$ matrix}\quad \Gamma \quad.
\ee
Generically, the freedom of basis change in im$P$ can be used to diagonalize
\be
\Gamma\ \to\ \textrm{diag}(\a_1,\a_2,\dots,\a_k) 
\qquad\textrm{with}\quad \a_i\in\C \quad,
\ee
so that we have
\be
a\,|T_i\> \= |T_i\>\,\a_i \qquad\Longrightarrow\qquad
|T_i\> \= |\a_i\> \ \equiv\ \e^{\a_i a^\+}\,|0\> \quad,
\ee
revealing the key role of coherent states.
We note that our solution depends on $k$~complex moduli parameters.
The ensueing BPS projector
\be \label{Palpha}
P_\a \= \sum_{i,j=1}^k 
|\a_i\> \,\bigl( \<\a_.|\a_.\> \bigr)^{-1}_{ij} \<\a_j|
\ee
generates a superposition of $k$~Gaussian lumps in the Moyal plane.
Besides the (inessential) choice of normalizations, 
only a residual permutation freedom remains in the solution
$|T\>=|\a\>:=\bigl(\,|\a_1\>,|\a_2\>,\dots,|\a_k\>\bigr)$. This corresponds 
to a relabelling of the lumps and emphasizes their bosonic character.
The general situation allows for coinciding values of some $\a_i\to\a$, 
which leaves $\Gamma$ in Jordan form. A Jordan block of size~$r$ yields
a sub-basis $\bigl\{\,|\a\>, a^\+|\a\>, \ldots, (a^\+)^{r-1}|\a\> \bigr\}$, 
whose span is obviously invariant under the action of~$a$. 
Clearly, the `fusion' of lumps smoothly produces lumps of higher `weight'. 
These observations determine the moduli space~${\cal M}_k$ as the 
$k$-th symmetrized power of the complex plane, i.e.
\be
{\cal M}_k \= \C^k/S_k \ \cong\ \C^k \quad,
\ee
which is a smooth K\"ahler manifold despite the coordinate singularities 
at the coincidence loci~\cite{goheaspr}.

\bigskip

\section{Soliton moduli-space action for a family of abelian sigma models}
\noindent
We formulate the action for the extended noncommutative abelian sigma model
in star-product language. For the group-valued field
\be
\phi \ \in\ \textrm{U}_\star(1) \quad,\qquad \textrm{i.e.}\quad
\phi\star\phi^\+ \= \unity \= \phi^\+\star\phi \quad,
\ee
we define the antihermitian composite flat gauge connection
\be
J\ :=\ \phi^\+\star\diff\phi \qquad\longrightarrow\qquad
F\ \equiv\ \diff J + J\wedge J \=0 \quad.
\ee
The action is a sum
\be \label{Sgamma}
S_\gamma \= S_2\ +\ \gamma\, S_3 \qquad\textrm{for}\quad \gamma\in[0,1]\quad,
\ee
where standard sigma-model term
\be
S_2 \= \sfrac12\int J\wedge *J
\ee
is formulated with a wedge product based on the star product and 
the Hodge star depending on the $1{+}2$ dimensional Minkowski metric~$\eta$. 
The WZW-like term
\be
S_3 \= -\sfrac13\int_0^1\!\!\int 
V\wedge\widetilde{J}\wedge\widetilde{J}\wedge\widetilde{J}
\ee
is an integral over $\R^{1,2}_\theta\times[0,1]$, with the extension
$\widetilde{J}$ interpolating along the interval~$[0,1]$ between 
$\widetilde{J}{=}0$ and $\widetilde{J}{=}J$. 
Furthermore, there appears the Lorentz-breaking constant one-form
\be
V \= \diff x \qquad\textrm{for}\quad z\=x+\im y \quad.
\ee
By varying $\gamma$, we get a family of actions connecting
the ordinary (non-chiral) sigma model (at $\gamma{=}0$) to 
the (chiral) Ward model (at $\gamma{=}1$), both based on $\textrm{U}_\star(1)$.

Introducing coordinates $(x^\mu)=(x^0,x^1,x^2)=(t,x,y)$ and subjecting
the field to a general coordinate transformation
$\delta\phi=\xi^\mu(x)\,\pa_\mu\phi$,
the action changes by
\be
\delta S_\gamma \= \int \bigl\{
\pa^{(\mu}\xi^{\nu)}\, T_{\mu\nu}\ \diff^2{z}\,\diff{t}\ +\ 
\gamma\,V(\xi)\,J\wedge J\wedge J \bigr\} \quad,
\ee
which, writing $J=J_\mu\diff x^\mu$, yields the standard energy-momentum tensor
\be
T_{\mu\nu} \= J_\mu\star J_\nu -\sfrac12\eta_{\mu\nu}\,
\eta^{\rho\sigma} J_\rho\star J_\sigma \quad.
\ee
The Lorentz group~SO(1,2) is broken to the $y$-boosts by the choice
of~$V$ and, independently, to the $xy$-rotations by the Moyal deformation, 
leaving nothing. Since $y$- and $t$-translations are unbroken~\footnote{
Under $x$-translations, 
$\delta S_\gamma\sim\gamma\int J\wedge J\wedge J=24\pi^2\gamma\,n$ 
with $n\in\pi_3(\cal G)=\Z$ mostly. 
However, $n=0$ for $\phi\in\textrm{Gr}(k,\cal H)$. }
the energy functional
\be
E \= \sfrac12\int\!\diff^2{z}\ \bigl\{ \pa_t\phi^\+\,\pa_t\phi + 
\pa_x\phi^\+\,\pa_x\phi + \pa_y\phi^\+\,\pa_y\phi \bigr\}
\ee
is conserved for all values of~$\gamma$.

Finally, the equation of motion reads
\be
\bal
0 &\= (\eta^{\mu\nu}+\gamma\,V_\rho\,\varepsilon^{\rho\mu\nu})\,
\partial_\mu(\phi^\+\star\partial_\nu\phi) \\[6pt]
&\= \partial_x(\phi^\+\star\partial_x\phi) + 
(1{-}\gamma)\,\partial_y(\phi^\+\star\partial_y\phi) -
(1{-}\gamma)\,\partial_t(\phi^\+\star\partial_t\phi) +
\gamma\,\partial_{y-t}(\phi^\+\star\partial_{y+t}\phi)
\eal
\ee
with the Minkowski metric $(\eta_{\mu\nu})=\textrm{diag}(-1,+1,+1)$
and the Levi-Civita tensor~$\varepsilon$, where $\varepsilon^{012}=1$. 

For the adiabatic approximation, we need to find the static multi-lump
solutions~$\ph(x,y)$. Since static configurations do not contribute to $S_3$, 
the energy $E$ reduces to $E_1$ in~(\ref{E1}). Hence,
the moduli space of static multi-lumps is the same for all $\gamma$,
namely ${\cal M}_k$ as derived in previous section.
Abbreviating the $k$ complex moduli by~$\a$, we denote the static $k$-lump
solution by $\ph(z,\zb;\a)$.
To extract the time dependence in the action, we rewrite the latter as
\be
S_\gamma[\phi] \= \int\!\diff{t}\,\diff^2{z}\ \bigl[ \sfrac12\dot\phi^2\ +\ 
C_\star(\phi,\phi')\,\dot\phi\ -\ W_\star(\phi,\phi') \bigr]
\qquad\textrm{with}\quad \phi'\equiv(\pa_z\phi,\pa_{\bar z}\phi) \quad.
\ee
Manton posits that slow soliton motion follows a geodesic of the static 
moduli space~${\cal M}_k$, i.e.
\be
\ph(t,z,\zb)\ \approx\ \ph(z,\zb;\a(t))\ =:\ \phi_\a \quad,
\ee
thus replacing dynamics for~$\ph(t,z,\zb)$ with dynamics for~$\a(t)$.
We are instructed to compute
\be
\bal
S_{\text{mod}}[\a] &\ :=\ S[\phi_\a] \\[6pt]
&\= \int\!\diff{t}\; \bigl[\,
\sfrac12 \{\smallint(\pa_\a\phi_\a)^2\}\, \dot\a^2 \ +\
\{\smallint C_\star(\phi_\a,\phi'_\a)\,\pa_\a\phi_\a\}\,\dot\a \ -\ 
\smallint W_\star(\phi_\a,\phi'_\a)\,\bigr] \\[6pt]
&\ =:\,\int\!\diff{t}\; \bigl[\,
\sfrac12 g_{\a\a}(\a)\, \dot\a^2\ +\ A_\a(\a)\, \dot\a\ -\ U(\a)\,\bigr] 
\eal
\ee
and read off the metric~$g$, magnetic field~$F=\diff{A}$ and
potential~$U$ on the moduli space.

To implement this program for the extended deformed abelian sigma model, 
we return to the operator formulation. 
Putting $\a_i\to\a_i(t)$ introduces $t$-dependence into
\be
|\a\> = \bigl(|\a_1\>,\dots,|\a_k\>\bigr) \quad\to\quad
P_\a = |\a\> \sfrac{1}{\<\a|\a\>} \<\a| \quad\to\quad
\phi_\a = \unity{-}2P_\a \quad\to\quad
\widehat{J} = 2\,[P_\a,\diff P_\a ] \quad.
\ee
Inserting the obtained $\widehat{J}(\a,\dot\a)$ 
into the action~(\ref{Sgamma}) we make two important observations.
Firstly, 
\be
S_3[\phi_\a] \ \sim\ \int\!\diff{t}\ \pa_t\ \tr_{\cal H} 
\bigl[\, (a+a^\+)\,P_\a \,\bigr] \=
\int\!\diff{t}\ \Bigl\{ \pa_{\a_i}\,\tr_{\cal H}
\bigl[\, (a+a^\+)\,P_\a \,\bigr] \Bigr\} \ \dot\a_i \ +\ c.c.\quad,
\ee
which reveals the magnetic potential to be exact, $A_i=\pa_{\a_i}\Omega$.
Thus, magnetic forcing is absent,\footnote{
The holonomy of $A=A_\a\diff\a$ may yet be nontrivial.}
and $S_{\text{mod}}$ is independent of~$\gamma$.
Secondly, we get
\be
S_2[\phi_\a] \= \int\!\diff{t}\ \tr_{\cal H}
\Bigl[\, \pi\theta\,\bigl|\dot\phi_\a\bigr|^2 \ -\ 
2\pi\,\bigl|[a,\phi_\a]\bigr|^2 \Bigr] \=
\int\!\diff{t}\ \Bigl[\,4\pi\theta\,\tr_{\cal H} \dot P_\a^2\ -\ 
E[\phi_\a]\Bigr] \quad.
\ee
Because $E[\phi_\a]=8\pi\,k$ is constant inside~${\cal M}_k$,
the second term yields an irrelevant constant potential~$U$ and can be dropped. 

As a result, $S_{\text{mod}}$ reduces to the kinetic part of $S_2[\phi_\a]$,
which simplifies to
\bea
S_{\text{mod}} \= 4\pi\theta\int\!\diff{t}\ \tr_{\cal H} \dot P_\a^2
&=& 8\pi\theta\int\!\diff{t}\ \tr_{\cal H}\ 
(\unity{-}P_\a)\,|\dot\a\>\,\<\a|\a\>^{-1} \<\dot\a| \\[6pt] \nonumber
&=& 8\pi\theta\int\!\diff{t}\ \tr_k\ 
\<\a|\a\>^{-1} \<\dot\a|\unity{-}P_\a|\dot\a\> \ =:\ 
\int\!\diff{t}\ \sum_{i,j=1}^k g_{\bar\imath j}\ \dot\ab_i\,\dot\a_j\quad,
\eea
where
\be
|\dot\a\>\ \equiv\ \pa_t|\a\>\= a^\+|\a\>\,\dot\Gamma
\qquad\textrm{with}\quad \Gamma=\textrm{diag}(\{\a_i\})\quad.
\ee
Hence, abbreviating $\pa_{\a_j}\equiv\pa_j$ and 
$\pa_{\bar\imath}\equiv\pa_{\ab_i}$, the metric on~${\cal M}_k$ is given by
\be
g_{\bar\imath j} \= 
8\pi\theta\ \tr_{\cal H}\ \pa_{\bar\imath} P_\a\,\pa_j P_\a \=
8\pi\theta\ \tr_k\ \<\a|\a\>^{-1}\,\pa_{\bar\imath}\Gamma^\+\,
\<\a|a\,(\unity{-}P_\a)\,a^\+|\a\>\, \pa_j \Gamma \quad.
\ee
With the shorthand
\be
M \= (M_{ij}) \ :=\ \bigl( \<\a|\a\>_{ij} \bigr) 
\= \bigl( \<\a_i|\a_j\> \bigr) \= \bigl(\!\e^{\bar\a_i\a_j} \bigr)
\ee
one computes
\be
\bal
S_{\text{mod}} &\= 8\pi\theta\int\!\diff{t}\ \tr_k\
\<\a|\a\>^{-1}\,\dot\Gamma^\+\,\<\a|a\,(\unity{-}P_\a)\,a^\+|\a\>\,
\dot\Gamma \\[6pt]
&\= 8\pi\theta\int\!\diff{t}\ \sum_{i,j=1}^k\
M^{-1}_{ji}\,\dot\ab_i\,\<\a_i|a\,(\unity{-}P_\a)\,a^\+|\a_j\>\,\dot\a_j\\[6pt]
&\= 8\pi\theta\int\!\diff{t}\ \sum_{i,j=1}^k\ M^{-1}_{ji}\,\Bigl\{
M_{ij} (1+\ab_i\a_j)\ -\ {\textstyle\sum_{m,n=1}^k} 
M_{im}\,\a_m\,M^{-1}_{mn}\,\ab_n\,M_{nj} \Bigr\} \ 
\dot\ab_i\,\dot\a_j \\[6pt]
&\= 8\pi\theta\int\!\diff{t}\ \sum_{i,j=1}^k\ M^{-1}_{ji}\,\Bigl\{
M\ +\ \Gamma^\+\,M\,\Gamma\ -\ M\,\Gamma\,M^{-1}\Gamma^\+\,M\Bigr\}_{ij}\ 
\dot\ab_i\,\dot\a_j \quad,
\eal
\ee
which reveals the hermitian metric~$(g_{\bar\imath j})$ on~${\cal M}_k$.
Using the identities
\be
\pa_j M \= \Gamma^\+\,M\,\pa_j\Gamma \und
\pa_{\bar\imath} M \= \pa_{\bar\imath}\Gamma^\+\,M\,\Gamma \quad,
\ee
it is straightforward to check that this metric is indeed K\"ahler
and derives from the K\"ahler potential
\be \label{Kahler}
K \= 8\pi\theta\,\ln\,\det\,M 
\= 8\pi\theta\,\ln\,\det\,\bigl(\<\a_i|\a_j\>\bigr)
\= 8\pi\theta\,\ln\,\det\,\bigl(\!\e^{\bar\a_i\a_j} \bigr) \quad.
\ee
This result agrees with the geometric intuition:
up to the prefactor of~$8\pi\theta$, the metric
$g_{\bar\imath j}=\pa_{\bar\imath}\pa_j K$ is the natural one on
the Grassmannian Gr($k,\cal H$).
It also has an interesting interpretation in terms of a system of
classical identical particles~\cite{haisleili}.

Global rotations and translations act as
\be
|\a_i\>\ \to\ |\!\e^{\!\im\vartheta}\a_i\>\=\e^{\!\im\vartheta\,a^\+a}\,|\a_i\>
\und
|\a_i\>\ \to\ |\a_i{+}\beta\> \= \e^{\beta\,a^\+}|\a_i\> \quad,
\ee
respectively, and shift the K\"ahler potential by a gauge transformation,
\be
K \ \to\ K \ +\ 8\pi\theta\,{\textstyle\sum_{i=1}^k} \bigl(
\bar\beta\a_i  + \beta\ab_i + \beta\bar\beta \bigr) \quad,
\ee
leaving the metric unchanged. 
Furthermore, (\ref{Kahler}) is invariant under permutations of the $\a_i$.
When passing to center-of-mass and barycentric coordinates
\be
s \= \sfrac1k\,{\textstyle\sum_{i=1}^k} \a_i \und w_i \= \a_i - s 
\qquad\textrm{such that}\quad {\textstyle\sum_{i=1}^k}w_i =0 \quad,
\ee
we get the decomposition
\be
K \= 8\pi\theta\,k\,|s|^2 \ +\ 
8\pi\theta\,\ln\,\det\,\bigl(\!\e^{\bar w_i w_j} \bigr) \quad,
\ee
which shows that the metric depends only on difference 
coordinates~$\a_i{-}\a_j$. One may also extract the diagonal (free) part via
\be
K\= 8\pi\theta\,{\textstyle\sum_{i=1}^k}\,|\a_i|^2 \ +\ 8\pi\theta\,\ln\,\det\,
\bigl(\!\e^{-\frac12|\a_i-\a_j|^2\ +\ \frac12(\ab_i\a_j-\ab_j\a_i)}\bigr)\quad.
\ee
{}From this expression it is easy to see a cluster decomposition property:
Upon splitting the moduli into two groups, $\{\a_i\}=\{\a'_\ell,\a''_m\}$,
and separating these to infinity, 
\be
\lim_{|\a'_\ell-\a''_m|\to\infty} K \bigl(\{\a_i\}\bigr) \=
K \bigl(\{\a'_\ell\}\bigr)\ +\ K \bigl(\{\a''_m\} \bigr) \quad.
\ee
In particular, an isolated single lump at~$\a_q$ asymptotically contributes
with $|\a_q|^2$ to~$K$.
Therefore, the moduli-space metric becomes flat for large mutual separations,
$|\a_i{-}\a_j|\to\infty$.

More interesting is the limit of coinciding lumps, say 
$\a_i\to\a$ for $i=q_1,\ldots,q_r$. Some lengthy algebra shows that then
\be \label{coincide}
K\ \to\ 8\pi\theta\sum_{q_\ell>q_m} \ln\,|\a_{q_\ell}{-}\a_{q_m}|^2 \ +\ K' 
\qquad\textrm{where still}\quad
K' \= \ln\,\det\,\bigl(\<T|T\>\bigr) \quad,
\ee
but after making inside $|T\>=\bigl(\,|\a_1\>,\ldots,|\a_k\>\bigr)$
the replacement
\be \label{Jordan}
\bigl\{\,|\a_{q_1}\>, |\a_{q_2}\>, \ldots, |\a_{q_r}\> \bigr\} \ \to\ 
\bigl\{\,|\a\>, a^\+|\a\>, \ldots, (a^\+)^{r-1}|\a\> \bigr\} \quad.
\ee
The coordinate singularity in (\ref{coincide}) can be removed by passing
to new coordinates, namely elementary symmetric polynomials in 
$\a_{q_\ell}{-}\a$, which correspond precisely to the new states 
in~(\ref{Jordan}). $K'$~produces the same metric as~$K$ but is smooth
at the coincidence locus. In the most general situation, $|T\>$ is composed
of various blocks like in~(\ref{Jordan}), of different sizes~$r$, 
but the formula for the smooth K\"ahler potential~$K'$ in~(\ref{coincide}) 
remains correct.

\bigskip

\section{Moduli-space trajectories for two-soliton scattering}
\noindent
For concreteness, let us display the simplest nontrivial case, i.e.~$k=2$.
The moduli space~${\cal M}_2$ of rank-two BPS projectors is parametrized
by $\{\a,\b\}\simeq\{\b,\a\}\in\C^2/S_2$.
Since the details have been given in~\cite{liroun}, we can be short here.

The static two-lump configuration is derived from~(\ref{Palpha}) as
\be \label{twolump}
\phi_{\a\b} \= \unity-2P_{\a\b} \= \unity\ -\ 2\,\frac{
|\a\>\<\b|\b\>\<\a|+|\b\>\<\a|\a\>\<\b|-|\a\>\<\a|\b\>\<\b|-|\b\>\<\b|\a\>\<\a|}
{\<\a|\a\>\<\b|\b\>-\<\a|\b\>\<\b|\a\>} \quad.
\ee
Writing $\a=s{+}w$ and $\b=s{-}w$ as well as $2w=:r\e^{\!\im\varphi}$,
the corresponding K\"ahler potential reads
\be
\bal
K&\= 8\pi\theta\,\ln\,\bigl(\!\e^{\a\ab+\b\bar\b}-\e^{\a\bar\b+\b\ab}\bigr)\!\!
&=&\ \ 8\pi\theta \bigl[ 2\,s\bar s\ +\ 
2\,w\bar w\ +\ \ln\,\bigl(1-\e^{-4w\bar w}\bigr) \bigr] \\[6pt]
&\= 8\pi\theta\,\ln\,\bigl(2\e^{2s\bar s}\sinh\sfrac{r^2}{2}\bigr) \!\! &=&\ \ 
8\pi\theta\bigl[2\,s\bar s\ +\ \sfrac12\,r^2\ +\ \ln\,\bigl(1-\e^{-r^2}\bigr)
\bigr] \quad,
\eal
\ee
with the limits
\be \label{Klimits}
K \=8\pi\theta\bigl[ 2s\bar s + \sfrac12 r^2 - \e^{-r^2} + O(\e^{-2r^2}) \bigr]
\und
K \= 8\pi\theta\bigl[ 2s\bar s + \ln r^2 + \sfrac{1}{24}r^4 + O(r^8) \bigr]
\quad.
\ee
It yields the metric
\be
\diff \ell^2 \= 16\pi\theta 
\bigl[ \diff{s}\diff{\bar s}\ +\ \Omega\,\diff{w}\diff{\bar w} \bigr]
\= 4\pi\theta \bigl[ 4\,\diff{s}\diff{\bar s}\ +\ 
\Omega(r^2)\,(\diff{r}^2 + r^2 \diff{\varphi}^2) \bigr]
\ee
with the conformal factor
\be \label{conf}
\Omega(r^2) \= 
\sfrac{1}{4\pi\theta}\,\pa_{r^2} \bigl( r^2\,\pa_{r^2} K \bigr) \=
\frac{1-2r^2\e^{-r^2}-\e^{-2r^2}}{(1-\e^{-r^2})^2} \=
\frac{\sinh r^2 - r^2}{\cosh r^2 - 1} \quad,
\ee
possessing the limits
\be
\Omega(r^2) \= 1 + (1{-}2r^2)\e^{-r^2} + O(\e^{-2r^2}) \und
\Omega(r^2) \= \sfrac13 r^2 - \sfrac{1}{90} r^6 + O(r^{10}) \quad.
\ee
\vskip-5mm
\begin{figure}[H]
\psfrag{$Omega$}{$\Omega$}
\psfrag{$r$}{$r$}
\begin{center}
\includegraphics[origin=ct,width=80mm]{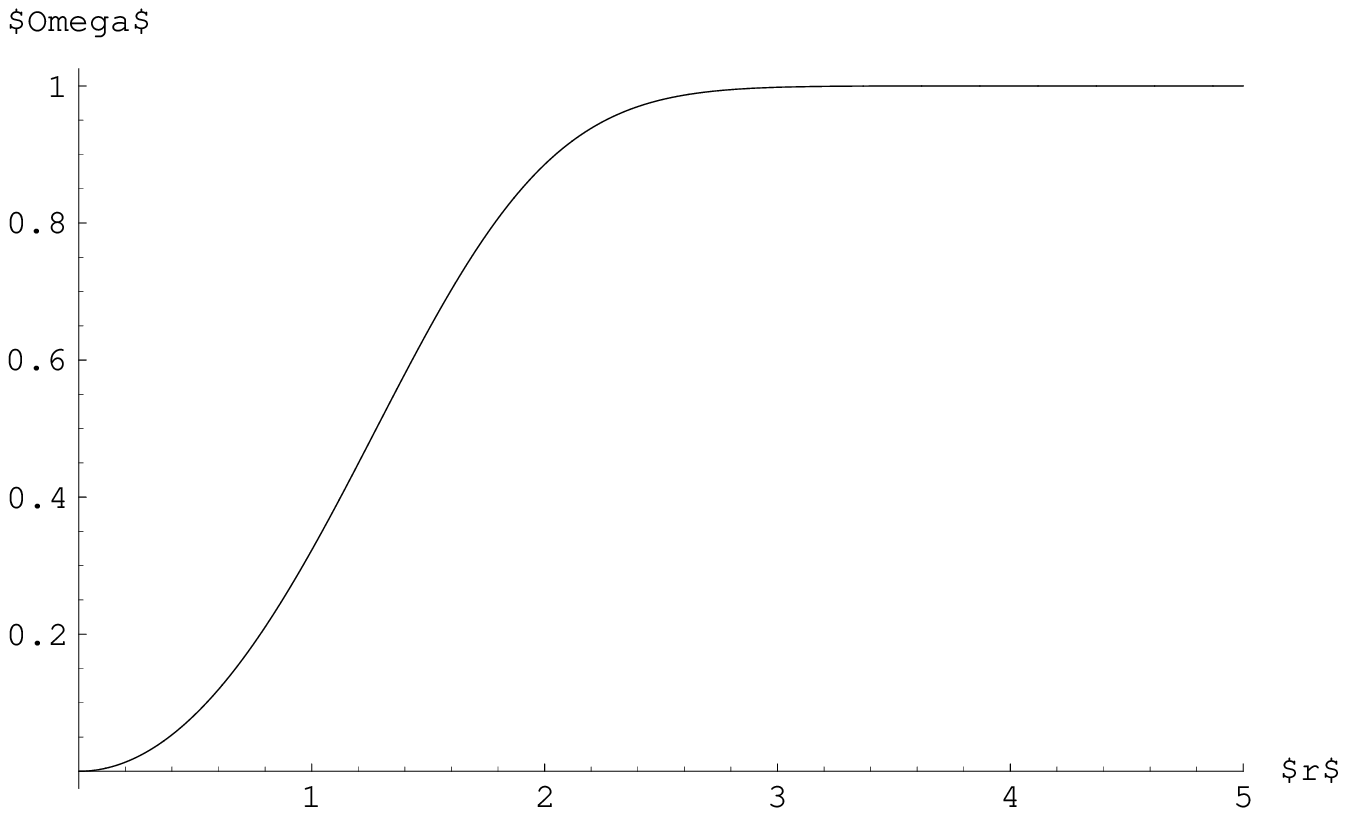}
\end{center}
%\caption{}
\end{figure}
\vskip-5mm 
Clearly, the metric becomes flat for $r\to\infty$ but develops a
conical singularity with an angle of~$4\pi$ at~$r=0$.
The latter is removed by passing to the symmetric coordinate
$\sigma=w^2$, in terms of which one finds
\be
\diff \ell^2 \= 16\pi\theta \Bigl[ \diff{s}\diff{\bar s}\ +\ 
\sfrac{\Omega(r^2\to4\sqrt{\sigma\bar\sigma})}{4\sqrt{\sigma\bar\sigma}}
\,\diff{\sigma}\diff{\bar\sigma} \Bigr]
\= 16\pi\theta \Bigl[ \diff{s}\diff{\bar s}\ +\ 
\bigl(\sfrac13-\sfrac{8}{45}\sigma\bar\sigma+O((\sigma\bar\sigma)^2)\bigr)
\,\diff{\sigma}\diff{\bar\sigma}\Bigr] \quad.
\ee
Due to the decoupling of the trivial center-of-mass dynamics,
${\cal M}_2=\C\times{\cal M}_\textrm{rel}$, 
with ${\cal M}_\textrm{rel}\simeq\C$ rotationally symmetric, 
asymptotically flat, and of positive curvature
$R=\frac{1}{4\pi\theta}[\frac{5}{4}-\frac{6}{175}r^4+O(r^8)]$.
Head-on scattering of two lumps corresponds to a single radial trajectory 
in~${\cal M}_\textrm{rel}$, which in the smooth coordinate~$\sigma$ 
must pass straight through the origin. 
In the `doubled coordinate'~$w=\sqrt{\sigma}$, we then see two straight 
trajectories with $90^\circ$ scattering off the singularity in the Moyal plane.

This picture persists for the scattering of two composite lumps, 
i.e.~lumps obtained by fusing
\be
\a_i \ \to\ \a \quad\textrm{for}\quad i=1,\ldots,r_1 \und
\a_{r_1+j}\ \to\ \b \quad\textrm{for}\quad j=1,\ldots,r_2 \quad.
\ee
The decoupling of the center-of-mass coordinate (now for $k=r_1{+}r_2$)
is achieved by writing
\be
\a \= s + r_2 w \und \b \= s - r_1 w \qquad\textrm{such that}\qquad
\a - \b \= (r_1{+}r_2)\,w \ =:\ r\e^{\!\im\varphi} \quad,
\ee
and one obtains
\bea
K&=&8\pi\theta\bigl[ (r_1{+}r_2) s\bar s\ +\ \sfrac{r_1 r_2}{r_1+r_2}\,r^2\ +\
\ln\,\bigl(1-{\cal P} \e^{-r^2} + O(\e^{-2r^2})\bigr) \bigr]  \\[6pt]
&\buildrel{r\to 0}\over\longrightarrow& 
8\pi\theta\bigl[ (r_1{+}r_2) s\bar s\ +\ 
c_0\ +\ c_1\ln r^2\ +\ c_2\,r^4\ +\ O(r^6) \bigr] \quad,
\eea
where ${\cal P}$ is a polynomial in~$r^2$ 
and $c_0$, $c_1$ and $c_2$ are constants.
As in (\ref{Klimits}), the absence of the $r^2$ term (and $c_2\neq0$)
leads to a conformal factor $\Omega\sim r^2$ for $r\to0$ and the same
conical singularity for any value of $r_1$ or $r_2$. Its remedy by 
employing the coordinate $\sigma=w^2$ demonstrates that 
the $90^\circ$~scattering angle is universal for head-on motion.
Only for more special situations with simultaneous head-on collision
of $k\,({>}2)$ solitons one will get $\frac\pi k$ scattering.

Let us return to the simple case of $r_1=r_2=1$ and drop the center-of mass
coordinate. The motion in ${\cal M}_\textrm{rel}$ is geodesic with conformal
factor~$\Omega(r^2)$ given in~(\ref{conf}). It conserves angular momentum
and energy,
\be \label{conserved}
l \= \Omega\,r^2\,\dot\phi \= v_\infty\,b \und 
e \= \sfrac12\,\Omega\,\dot r^2 + \frac{l^2}{2\Omega r^2} 
\= \sfrac12\,v_\infty^2 \quad,
\ee
respectively, with the asymptotic speed~$v_\infty$ and the impact parameter
\be
b \= l/\sqrt{2e} \= r_\textrm{min}\,\sqrt{\Omega(r_\textrm{min})} \quad.
\ee
Hence, the trajectory is given by
\be
\frac{\diff r}{\diff \phi} \= \frac{r^2}{b}\,\sqrt{\Omega-b^2/r^2} \quad,
\ee
and we obtain the scattering angle
\be
\Theta(b) \= \pi\ -\ 2 \int_{r_\textrm{min}}^\infty 
\frac{b\ \diff{r}}{r^2\,\sqrt{\Omega-b^2/r^2}} \quad,
\ee
which varies between $0$ (for $b\to\infty$ and $\Omega\to1$) and 
$\frac\pi2$ (for $b\to0$ and $\Omega\to\frac13r^2$).
Therefore, if we fix~$b$ and vary~$v_\infty$, the trajectory is unchanged.
The total energy of the $k{=}2$ system is
\be \label{Emod}
E_\textrm{mod} \ :=\ E[\phi_\a] \= 16 \pi + 8 \pi\theta\,e 
\= 16 \pi + 4 \pi\theta v_\infty^2 \quad.
\ee
Good agreement with the full field-theory dynamics is expected
only for small values of~$v_\infty$. For the cases 
$r_1=1(\textrm{triangles}),2(\textrm{diamonds}),4(\textrm{boxes})$ and $r_2=1$
the function $\Theta(b)$ is plotted below.
\vskip-5mm
\begin{figure}[H]
\psfrag{th}{$\Theta$}
\psfrag{b}{$b$}
\begin{center}
\includegraphics[origin=ct,width=80mm]{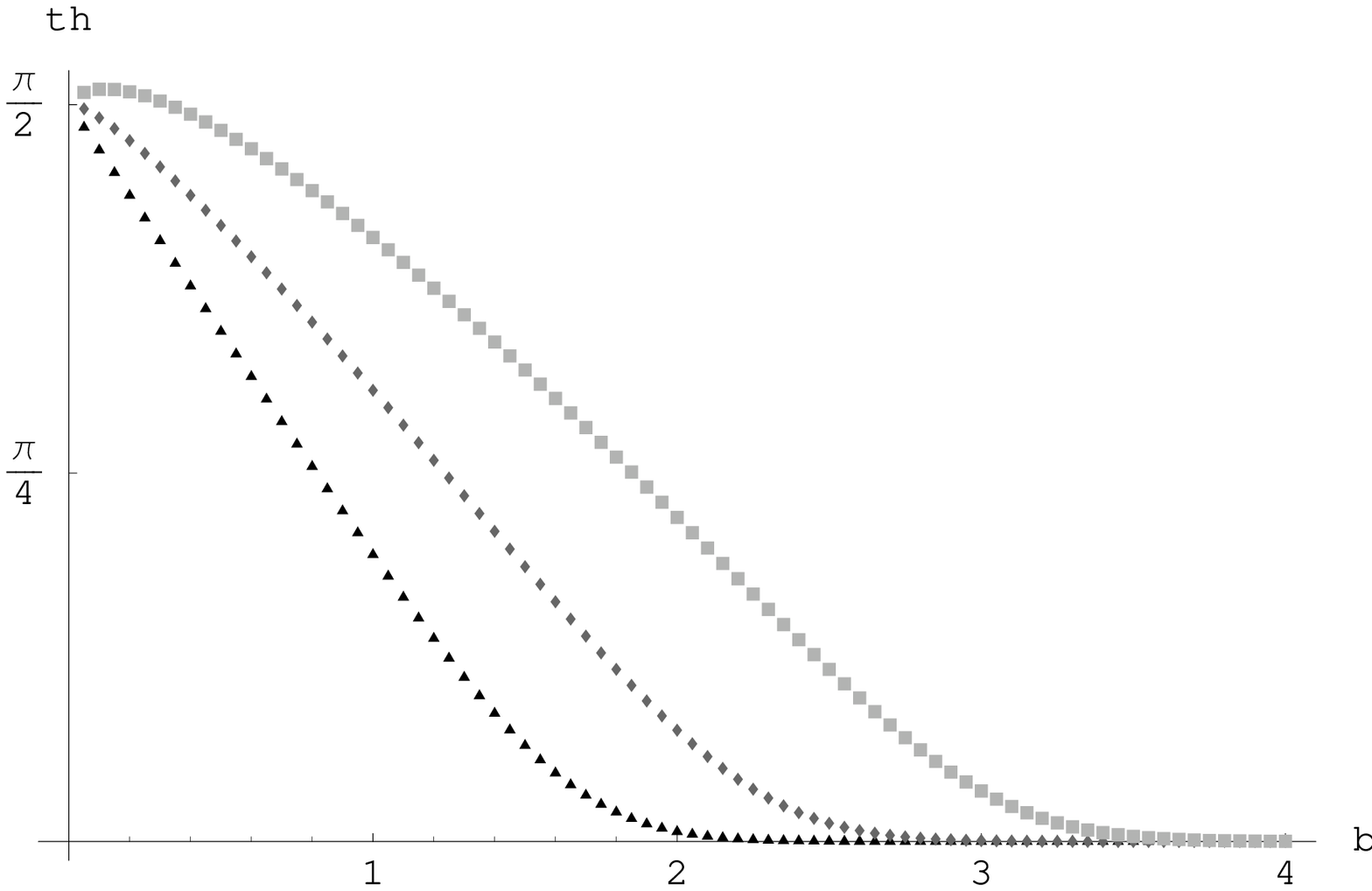}
\end{center}
%\caption{}
\end{figure}
\vskip-5mm

\bigskip

\section{Comparison with time-dependent field-theory solutions}
\noindent
According to the general arguments about the adiabatic approximation,
the moduli-space dynamics described in the previous section should apply
to the whole family~$S_\gamma$ of actions in~(\ref{Sgamma}).
To test the quality of the approach, one would like to compare the
moduli-space scattering trajectories with the time evolution of the
energy-density maxima of the corresponding classical field configurations.
Since widely separated lumps roam essentially independently of each other,
we already know the large-time asymptotics: a (multiplicative) superposition 
of several one-soliton configurations of the form $\phi_\a=\unity{-}2P_\a$,
after applying individual translations and boosts.

To simplify the discussion, let us consider just two lumps of rank one each,
i.e.~combine two copies of $\phi_0=\unity{-}2|0\>\!\<0|$.
For large (positive or negative) times we then must have
\be \label{asymp}
\bal
\phi(t\to\pm\infty) &\ \simeq\ 
\bigl(\unity -(1{-}\e^{\!\im\delta_1})\,U_1\,|0\>\!\<0|\,U_1^\+ \bigr)\,
\bigl(\unity -(1{-}\e^{\!\im\delta_2})\,U_2\,|0\>\!\<0|\,U_2^\+ \bigr) \\[6pt]
&\ \simeq\ \unity\ -\ (1{-}\e^{\!\im\delta_1})\,U_1\,|0\>\!\<0|\,U_1^\+ 
\ -\ (1{-}\e^{\!\im\delta_2})\,U_2\,|0\>\!\<0|\,U_2^\+ \quad,
\eal
\ee
where 
\be
U_i \= U(\vec{v}_i^{\,\pm},\vec{r}_i^{\,\pm},t) 
\qquad\textrm{for}\quad i=1,2 \quad\textrm{and}\quad t \to \pm\infty
\ee
are unitary transformations implementing translations by $\vec{r}_i^{\,\pm}$
and boosts with velocities~$\vec{v}_i^{\,\pm}$ in the Moyal plane.
Note that (time-dependent) solitons need no longer be Grassmannian,
and so we must allow in~(\ref{asymp}) for the slightly more general prefactors
with velocity-dependent phases~$\delta_i$~\cite{lepo1,lepo2}. 
If the scattering angle differs from~$\pi$, i.e.~if nontrivial scattering 
occurs, then the late velocities $\vec{v}_i^{\,+}$ must differ from the 
early ones~$\vec{v}_i^{\,-}$.

Outside the value $\gamma=1$, 
the solitons are affected by each other's presence,
and no integrability protects them from shrinking and decay.
Yet, in cases where their lifetime is sufficiently long 
the configurations~(\ref{asymp}) can still be approached 
for not too large times, and scattering data are viable.
In the absence of exact time-dependent solutions, however,
numerical investigations are needed for confirmation. 
To the author's knowledge, computer analysis has been applied only 
in the commutative case ($\theta{=}0$) for $\gamma{=}0$ and $\cal G=$~O(3),
where it established the universality of $90^\circ$ head-on two-soliton
scattering~\cite{zak}.
In the noncommutative realm, the equation of motion for $\cal G=$~U(1) 
and $\gamma{=}0$ to solve in the operator formulation is
\be
\bal
0 &\= \theta\,\pa_t (\phi^\+ \pa_t\phi) \ +\ 
\bigl[ a\,,\,\phi^\+[a^\+,\,\phi]\bigr]\ +\
\bigl[ a^\+,\,\phi^\+[a\,,\,\phi]\bigr] \\[6pt]
  &\= \theta\,\pa_t (\phi^\+ \pa_t\phi) \ +\
\phi^\+\,\bigl[a\,,\,[a^\+,\,\phi]\bigr]\ -\
\bigl[a\,,\,[a^\+,\,\phi^\+]\bigr] \,\phi \quad.
\eal
\ee
With $\phi\in\textrm{U}(\cal H)$ viewed as an infinite-size 
matrix~$(\phi_{mn})$ in the Fock-space basis~(\ref{nbasis}), it reads
\be
\bal
\theta\,\pa_t (\phi^*_{nm} \pa_t\phi_{n\ell}) \=
(m{-}\ell)\,\phi^*_{nm} \phi_{n\ell} \ &+\
\sqrt{(n{+}1)(\ell+1)}\,\phi^*_{nm} \phi_{n+1\;\ell+1} \ +\
\sqrt{n\,\ell}\,\phi^*_{nm} \phi_{n-1\;\ell-1} \\[6pt] &-\
\sqrt{(n{+}1)(m{+}1)}\,\phi^*_{n+1\,m+1} \phi_{n\ell} \ -\
\sqrt{nm}\,\phi^*_{n-1\,m-1} \phi_{n\ell} \quad.
\eal
\ee
It would be interesting to analyze this coupled initial-value problem 
numerically.

For $\gamma=1$, the situation is entirely different since exact multi-soliton 
solutions are available~\cite{lepo1,lepo2,wolf}. 
As a warm-up, consider the generic one-soliton configuration,
\be \label{onesol}
\phi_1(t) \= \bigl(\unity - \widetilde P(t)\bigr) +   
\sfrac{\mu}{\bar\mu}\,\widetilde P(t) \=
\unity\ -\ (1{-}\sfrac{\mu}{\bar\mu}) \widetilde P(t)
\qquad\textrm{with}\qquad
\widetilde P(t) \= U(\mu,t)\,P_\a\,U(\mu,t)^\+ \quad,
\ee
where $U(\mu,t)$ is the unitary transformation effecting a boost with velocity
\be \label{vel}
\vec{v} \ \equiv\ (v_x,v_y) \= -\Bigl(
\frac{\mu+\bar\mu}{\mu\bar\mu+1}\ ,\ \frac{\mu\bar\mu-1}{\mu\bar\mu+1}\Bigr)
\qquad\Longleftrightarrow\qquad
\mu \= -\frac{v_x+\im\sqrt{1{-}\vec{v}^2}}{1-v_y}\ \in\C\setminus\R \quad.
\ee
The energy of this configuration is found to be
\be
E[\phi_1] \ \equiv\ E(\vec{v}) \= 8\pi\,\frac{\sqrt{1-\vec{v}^2}}{1-v_y^2}
\= 8\pi\,(1-\sfrac12 v_x^2 +\sfrac12 v_y^2 + \ldots) \quad.
\ee

The exact two-soliton solution of rank two in the Ward model reads~\cite{lepo1}
\be \label{twosol}
\phi_2(t) \= \unity\ -\ \frac{
(1{-}\sfrac{\mu_1}{\bar\mu_1}) |1\>\!\<2|2\>\!\<1| +
(1{-}\sfrac{\mu_2}{\bar\mu_2}) |2\>\!\<1|1\>\!\<2| -
\mu(1{-}\sfrac{\mu_2}{\bar\mu_1}) |1\>\!\<1|2\>\!\<2| -
\mu(1{-}\sfrac{\mu_1}{\bar\mu_2}) |2\>\!\<2|1\>\!\<1|}{
\<1|1\>\!\<2|2\> - \mu \<1|2\>\!\<2|1\>}
\ee
with
\be
\mu \= \sfrac{(\mu_1-\bar\mu_1)(\mu_2-\bar\mu_2)}
             {(\mu_1-\bar\mu_2)(\mu_2-\bar\mu_1)}  \und
|\,i\,\> \= U(\mu_i,t)\,|\a_i\> \quad\textrm{for}\quad i=1,2 \quad.
\ee
Here, $\mu_i$ parametrize the (constant) velocities of the two solitons
like in~(\ref{vel}), and $\a_i$ are their positions at~$t{=}0$. 
The unitary transformations~$U$ boost the vacuum state, 
and so the time-dependent states~$|\,i\,\>$ are just moving-frame vacua 
for the two lumps. It is easy to verify that
in the static limit $\mu_i\to-\im$ the configuration~(\ref{twosol}) tends
to the static solution~(\ref{twolump}) as long as $\a_1\neq\a_2$.
For large times, the overlap~$\<1|2\>$ dies away, and indeed 
the form~(\ref{asymp}) is attained. However, we see that
$\vec{v}_i^{\,+}=\vec{v}_i^{\,-}=\vec{v}_i$ since the velocities do not change,
and thus there is no scattering!
This is also evinced by a no-force property of Ward solitons,
borne out by their energy additivity:
\be
E[\phi_2] \= E(\vec{v}_1) \ +\ E(\vec{v}_2) \quad.
\ee

Even for small velocities, the energy density of the solution (\ref{twosol}) 
does not follow the moduli-space dynamics of the previous section (except 
of course at very large impact parameter where the scattering disappears).
Therefore, we should look for other classes of exact two-soliton solutions.
Recently it has been established~\cite{daiteruhl} for the commutative case that 
all Ward model multi-solitons are obtained from the one-soliton configurations
(\ref{onesol}) by dressing and fusing operations.\footnote{
This result presumably extends to the noncommutative case.}
The energy is additive under dressing and unchanged under fusing. 
In fact, (\ref{twosol})~was constructed by dressing (\ref{onesol}) with a copy.

For comparison with the $k{=}2$~case of the previous section,
it remains to consider fusing the two-soliton~(\ref{twosol}). This is achieved
by putting $\a_1=\a_2=\a$ and sending both velocities to~zero.\footnote{
The more general situation of merely equal velocities is related by boosting
the center of mass.}
In this limit a new type of time dependence emerges. Putting in~(\ref{twosol})
\be
\mu_1 \= -\im + \epsilon \und \mu_2 \= -\im - \epsilon 
\qquad\textrm{with}\quad \C\ni\epsilon\to 0 
\ee
and observing that
\be
U(-\im\!{\pm}\epsilon,t)\,|\a\> \= \e^{-|\epsilon|^2\,t^2/4\theta}\,
\e^{\pm\a\,\epsilon\,t/\sqrt{2\theta}}\, \bigl( 1\ \mp\ 
\sfrac{\bar\epsilon t}{\sqrt{2\theta}}\,a^\+\ +\ O(\epsilon^2)\bigr) 
|\a\> \quad,
\ee
we learn that any time dependence comes in the combination
of~$\epsilon t/\sqrt{2\theta}$. It is crucial to observe that the limits
$\epsilon\to0$ and $|t|\to\infty$ do not commute, and so the asymptotic
behavior of~(\ref{twosol}) is modified under fusing. The result is
\be \label{fuseansatz}
\widetilde\phi_2(t)\ :=\ \lim_{\epsilon\to0}\phi_2(t) \=
\Bigl( \unity -2\sfrac{|\a\>\!\<\a|}{\<\a|\a\>} \Bigr)\,
\Bigl( \unity -2\sfrac{|\at\>\!\<\at|}{\<\at|\at\>} \Bigr)
\quad,
\ee
where the time dependence hides in
\be \label{atilde}
|\at\>\=|\a\>\ -\ \im\,t\,\sqrt{\sfrac2\theta}\;|\a^\perp\!\>
\qquad\textrm{with}\qquad |\a^\perp\!\> \= (a^\+ -\ab)\,|\a\> 
\ee
being orthogonal to~$|\a\>$. More explicitly,
\be \label{fusesol}
\widetilde\phi_2(t) \= \unity\ -\ \frac{2}{\theta+2t^2} \ \Bigl\{
2\,t^2\, \bigl( |\a\>\!\<\a| + |\a^\perp\!\>\!\<\a^\perp\!| \bigr) \ -\
\im t\,\sqrt{2\theta}\,\bigl(|\a\>\!\<\a^\perp\!| + |\a^\perp\!\>\!\<\a|\bigr)
\Bigr\} \quad,
\ee
which at $t{=}0$ momentarily degenerates to $\phi=\unity$.

This solution can also be constructed directly by the dressing method,
starting from the ansatz~(\ref{fuseansatz}) with an unknown state~$|\at\>$. 
In this way one arrives at the conditions~\cite{lepo2}
\be
a\,|\at\> + \bigl[ a\,,\sfrac{|\a\>\!\<\a|}{\<\a|\a\>} \bigr] |\at\>
\= |\at\>\,Z_1  \und
\pa_t\,|\at\> + \im\sqrt{\sfrac{2}{\theta}}\,
\bigl[a^\+,\sfrac{|\a\>\!\<\a|}{\<\a|\a\>}\bigr] |\at\> 
\= |\at\>\,Z_2 \quad,
\ee
where $Z_1$ and $Z_2$ are functions of~$t$ to be determined.
We read off that $Z_1=\a$ and fix the (inessential) normalization 
such that~$Z_2=0$. It is not hard then to recover~(\ref{atilde}) as the
general solution indeed.

Putting $\a{=}0$ for simplicity,
the energy density of~(\ref{fusesol}) is readily computed to be~\cite{lepo2}
\bea \label{edensity}
{\cal E} &=& \frac{4\theta}{(\theta{+}2t^2)^2}\ \Bigl\{ 
\bigl( |0\>\!\<0|+|1\>\!\<1| \bigr)\ +\ 
\frac{2t^2}{\theta} \bigl( 2\,|0\>\!\<0|+|1\>\!\<1|+|2\>\!\<2| \bigr)\ +\
\frac{4t^4}{\theta^2} \bigl( |1\>\!\<1|+|2\>\!\<2| \bigr) \\ 
\nonumber & & \qquad\ -\ 
\frac{2t^2}{\theta}\bigl(\sfrac1{\sqrt2}|2\>\!\<0|+\sfrac1{\sqrt2}|0\>\!\<2|
\bigr)\ -\ \im
\frac{2^{3/2}t^3}{\theta^{3/2}}\bigl( |1\>\!\<0|-|0\>\!\<1| +
\sfrac1{\sqrt2}|2\>\!\<1| - \sfrac1{\sqrt2}|1\>\!\<2| \bigr)\Bigr\} \quad,
\eea
with $E[\widetilde\phi_2]=2\pi\theta\,\tr{\cal E}=16\pi$ as should be.
Matching with $E_\textrm{mod}$ in~(\ref{Emod}) enforces $v_\infty=0$
which, however, does not restrict~$b$ in any way.
Employing the Moyal-Weyl correspondence, the energy density 
in the Moyal plane takes the form~\cite{lepo2}
\be
{\cal E}_\star\=\frac{16\,\e^{-r^2/\theta}}{\theta\,(1+2t^2/\theta)^2}\ \Bigl\{
\frac{r^2}{\theta}\ +\ 
\Bigl(1-\frac{r^2}{\theta}+\frac{r^4}{\theta^2}\Bigr)\,\frac{2t^2}{\theta}\ +\
\Bigl(-\frac{r^2}{\theta}+\frac{r^4}{\theta^2}\Bigr)\,\frac{4t^4}{\theta^2}\ -\
\Bigl(\frac{x^2}{\theta}-\frac{y^2}{\theta}\Bigr)\,\frac{2t^2}{\theta}\ -\
\frac{4y r^2 t^3}{\theta^3} \Bigr\} \quad.
\ee
Unfortunately, this energy distribution is invariant under spacetime inversion 
and has a ring-like structure in the Moyal plane, localized at the origin
like a bound state.
Hence, we do not find any scattering solutions with $k{=}2$ in the 
noncommutative U(1) Ward model.\footnote{
We generalized the ansatz~(\ref{fuseansatz}) by relaxing the first
projector, allowing for higher-rank projectors and admitting time-dependent
coefficients, all without success.}

Interestingly, the nonabelian Ward model is very different in this regard
because of its larger moduli space: Fusing the U(2) two-soliton solution 
also features ring-like configurations but also admits moduli choices 
which produce genuine $90^\circ$~scattering, 
in the commutative~\cite{ward3,ioa,anand,ioazak} 
as well as in the noncommutative~\cite{lepo2} case.
The corresponding moduli-space approximation was recently considered in
\cite{dunman} and \cite{finy}, respectively. For $\theta{=}0$, it seems to 
agree with the analytical and numerical field-theory results obtained earlier. 

In summary, their moduli-space motion does not approximate the extended 
abelian sigma-model soliton scattering in the Moyal plane equally well 
for all values of the family parameter~$\gamma$.
Numerical analysis is needed to make the case for~$\gamma{<}1$.
For the integrable value $\gamma{=}1$ (the abelian Ward model), however,
we are curiously lacking the field-theory dynamics which the moduli-space
kinematics is supposed to mimic.\footnote{
This also applies to the no-scattering solutions of the nonabelian models.}    
It is therefore conceivable that in this case,
even for arbitrary small velocities, the soliton scattering takes place far
away from their moduli space, if it occurs at all! Certainly, the known
no-scattering multi-solitons are not seen in the moduli space, 
which challenges Manton's paradigm. In part responsible for this failure
seems to be the absence of magnetic forcing in the moduli space in contrast 
to the crucial importance of the WZW-like action term for integrability.
Perhaps a numerical study can help to answer this conundrum.

\bigskip

\noindent
{\bf Acknowledgements}

\noindent
The authors are grateful to Andrei V.~Domrin and Alexander D.~Popov 
for discussions and reading the manuscript. This work was
partially supported by the Deutsche Forschungsgemeinschaft.

\bigskip
%\newpage

\end{document}